\documentclass[11pt]{article}

\usepackage[margin=1in]{geometry}
\usepackage{amsmath,amssymb,booktabs,graphicx,url,hyperref}
\usepackage[T1]{fontenc}
\usepackage[utf8]{inputenc}
\usepackage{microtype}
\usepackage{enumitem}
\usepackage{float}
\usepackage{multirow}
\usepackage{array}
\usepackage[numbers,sort&compress]{natbib}

\hypersetup{
    colorlinks=true,
    linkcolor=blue,
    citecolor=blue,
    urlcolor=blue
}

\title{Acoustic Biomarkers for Alzheimer\textquotesingle s Disease Detection from Spontaneous Speech}

\author{
  Anh L Tran\\
  Arjee Javellana Restar\\
  Bereket Kindo\\
  Kentro, Mclean, VA, US
}

\date{\today}

\begin{document}
\maketitle

\begin{abstract}
Alzheimer\textquotesingle s disease (AD) remains a major public health challenge, and early identification is crucial for timely intervention and care planning. Recent work has shown that speech-based digital biomarkers can capture subtle changes in vocal production, prosody, and timing that often precede or accompany cognitive decline. In this study, we investigate whether acoustic features extracted from spontaneous speech can reliably distinguish individuals with AD from healthy controls. Using a structured speech dataset derived from the Cookie Theft picture-description task, we extract clinically informative low-level descriptors spanning prosodic, spectral, and temporal domains, and evaluate a range of machine learning classifiers. Our results indicate that acoustic biomarkers carry strong discriminative value, with the best-performing models achieving clinically meaningful separation between diagnostic groups. These findings support the use of non-invasive, low-cost audio-based screening tools and motivate further work on multimodal and longitudinal assessment frameworks for dementia risk detection.
\end{abstract}


\section{Introduction}

The early detection of Alzheimer\textquotesingle s disease (AD) remains one of the central challenges in geriatric medicine and cognitive neuroscience. Traditional diagnostic pipelines rely on clinical interviews, neuropsychological testing, and imaging-based assessments, which are expensive, time-consuming, and often inaccessible for routine screening. As a result, there is growing interest in digital biomarkers that can be collected passively and analyzed at scale. Speech is particularly attractive because it is non-invasive, inexpensive to capture, and is highly sensitive to changes in cognition, executive function, and motor control \citep{eyben2016gemaps,haider2020assessment,luz2021speech}.

A large body of literature has shown that AD is associated with measurable changes in language production, including impaired lexical access, reduced speech fluency, and altered rhythm and prosody. These changes can be captured using acoustic features such as fundamental frequency variability, formant dynamics, pauses, spectral tilt, and temporal structure. In recent years, computational paralinguistics has emerged as a promising approach for deriving objective biomarkers from audio recordings without requiring manual transcription or extensive human annotation.

In this work, we examine the utility of acoustic-only features for AD detection in spontaneous speech. The goal is not to replace clinical diagnosis, but to evaluate whether audio-derived measurements can contribute useful, explainable, and scalable signals for clinical decision support. We consider a dataset consisting of speech recordings collected during a standardized picture-description task and train machine learning models to separate patients with AD from controls. By focusing on interpretable acoustic descriptors and robust classification pipelines, we aim to bridge the gap between signal processing research and clinically relevant screening methods.

The remainder of this article is organized as follows. We review prior work on acoustic and speech-based dementia detection, describe the dataset and feature extraction procedure, summarize the modeling strategy, and report experimental results. We conclude with a discussion of strengths, limitations, and avenues for future multimodal and longitudinal extensions.

\section{Related work}

Speech and language have long been recognized as useful indicators of cognitive decline. Several studies have shown that dementia is associated with reduced verbal fluency, increased hesitation, longer pauses, and altered prosodic patterns. More recently, computational analyses of spontaneous speech have demonstrated that acoustic characteristics can help distinguish AD from healthy aging, even in the absence of linguistic transcription.

The Geneva Minimalistic Acoustic Parameter Set (GeMAPS) has become a widely used feature framework in voice research and affective computing, and several of its descriptors are strongly related to prosodic and spectral changes that are clinically relevant in dementia \citep{eyben2016gemaps}. In general, frequency-related parameters (e.g., pitch and formant trajectories), amplitude-related parameters (e.g., loudness, energy), and spectral descriptors (e.g., MFCCs, spectral slope, HNR) have all been linked to impaired vocal control and reduced speech efficiency in neurodegenerative disease. This motivates the use of compact yet physiologically meaningful acoustic features that can be extracted reliably from short, real-world audio samples.

A particularly relevant line of work has focused on the Pitt corpus and similar Cookie Theft picture-description datasets. These studies often combine linguistic and acoustic information to improve classification, yet several works highlight the predictive value of purely acoustic features \citep{haider2020assessment,deng2022alzheimer,fraser2016linguistic}. In parallel, a broader corpus of dementia-speech studies has demonstrated that features such as articulation rate, pause duration, vocal stability, and cepstral dynamics are repeatedly associated with cognitive impairment. Together, these findings suggest that AD signatures are not limited to lexical or semantic content; they are also expressed in the physical and temporal properties of speech.

The present study contributes to this literature by evaluating a feature-driven acoustic modeling pipeline designed for spontaneous speech. Unlike approaches that rely heavily on manually transcribed text, we focus on signal-derived descriptors and interpretable classification models. This reflects the practical need for scalable screening tools that do not require expensive annotation pipelines or expert transcription workflow.

\section{Materials and Methods}

\subsection{Dataset}

The study uses a curated dataset of spontaneous speech recordings collected during the Boston Cookie Theft picture description task. This task is well established in neuropsychological assessment and provides a standardized context in which participants describe a complex scene. Each recording is labeled as normal cognition, mild cognitive impairment (MCI), or Alzheimer\textquotesingle s disease and related dementias (ADRD). The corpus is suitable for evaluating whether speech-derived measures can discriminate among these clinical groups.

\subsection{Preprocessing and feature extraction}

Audio recordings were standardized before analysis to reduce differences in recording quality and amplitude. This included resampling and amplitude normalization where appropriate, along with inspection for poor-quality segments or silent recordings. The acoustic feature extraction pipeline then computed a broad set of low-level descriptors from each speech sample.

The feature space included:
\begin{itemize}[leftmargin=1.5em]
    \item prosodic and pitch-related measures, including F0 statistics, pitch variability, and contour stability;
    \item energy and loudness features, including intensity, loudness peaks, and amplitude modulation;
    \item spectral descriptors, including MFCCs, spectral slope, and formant-related features;
    \item temporal structure measures, including pause statistics, voiced/unvoiced segment lengths, and speech rate proxies;
    \item variability and distributional descriptors summarizing frame-level dynamics over the recording.
\end{itemize}

This feature set is consistent with the acoustic literature on dementia detection and with established paralinguistic frameworks such as GeMAPS and eGeMAPS, which have been shown to capture important physiological and behavioral voice patterns linked to cognitive decline.

\subsection{Modeling framework}

We evaluated supervised models on several representations of the same recordings: statistically aggregated acoustic features, non-aggregated sequential acoustic features, textual transcriptions, and raw-audio embeddings. Aggregated acoustic features were classified with logistic regression, decision trees, random forests, and gradient-boosted ensembles (XGBoost, LightGBM, and CatBoost). Sequential features were modeled with an LSTM and a bidirectional LSTM. Transcriptions were represented with TF-IDF, Word2Vec, and a fine-tuned ALBERT model, and raw audio was encoded with Whisper. Stacked ensembles then combined complementary predictors. The primary metric was multiclass log loss, so that lower values indicate predicted probabilities closer to the true control, MCI, and ADRD labels. For interpretation, SHAP values were computed on the best aggregated-acoustic model.

The objective was to test whether a compact acoustic profile separates clinical groups, and whether linguistic and audio-embedding models add complementary signal, while keeping the pipeline reproducible.

\section{Experiments and Results}

\begin{figure}[H]
\centering
\includegraphics[width=\textwidth]{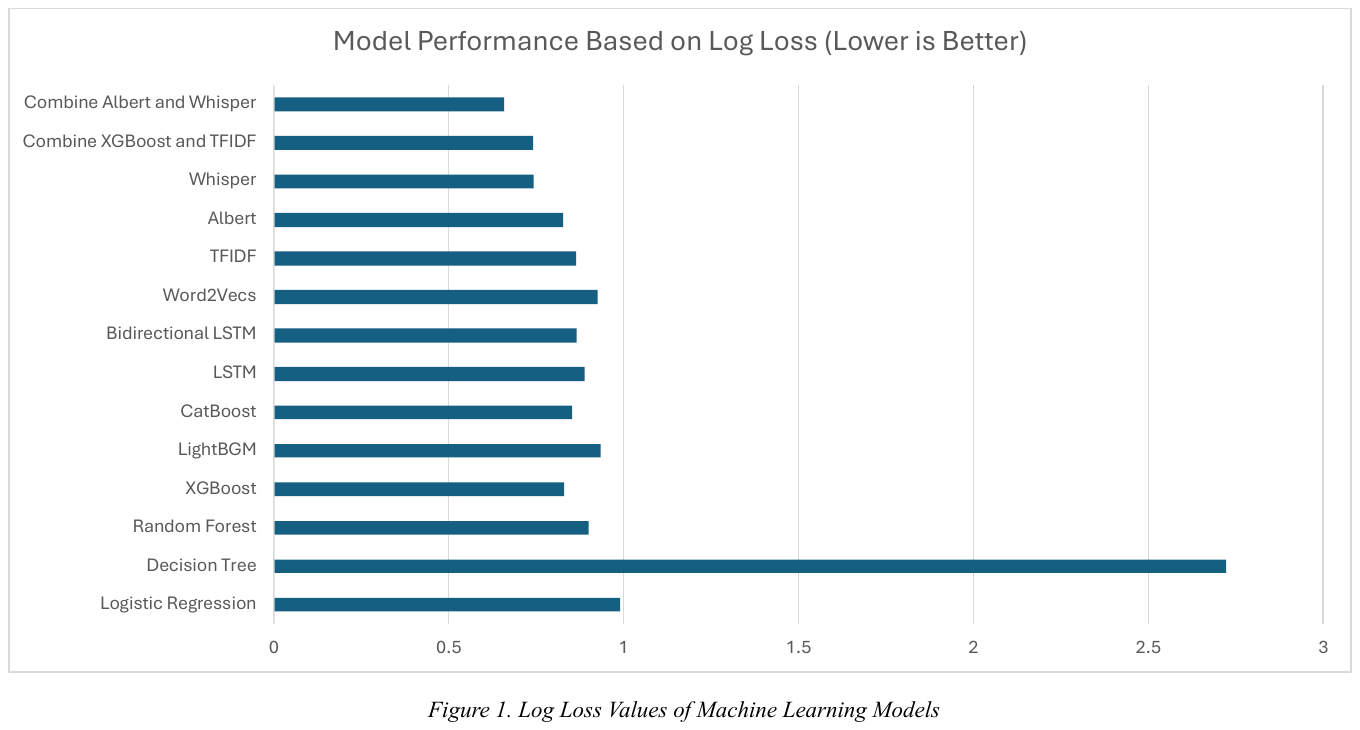}
\caption{Log loss of each model. Lower is better. Acoustic, sequential, transcription, audio-embedding, and ensemble models are shown together.}
\label{fig:logloss}
\end{figure}

Models were compared with multiclass log loss (Figure~\ref{fig:logloss}). Lower values indicate predicted probabilities closer to the true labels. Dementia-related speech changes are subtle and heterogeneous, so the spread across model families is as informative as the best single score.

Among classifiers trained on aggregated acoustic features, XGBoost had the lowest log loss (0.8299), followed by CatBoost (0.8530) and random forest (0.8989). Logistic regression was higher (0.9900). The decision tree was an outlier (2.7220), consistent with overfitting on this feature set. On non-aggregated sequential features, the bidirectional LSTM (0.8646) outperformed the unidirectional LSTM (0.8879). Among transcription models, ALBERT (0.8270) beat TF-IDF (0.8634) and Word2Vec (0.9251). Whisper, which embeds the waveform directly, reached 0.7430, the best score of any single model. Ensembles improved further: XGBoost stacked with TF-IDF reached 0.7404, and ALBERT stacked with Whisper reached 0.6580, the lowest log loss in the comparison.

\begin{figure}[H]
\centering
\includegraphics[width=\textwidth]{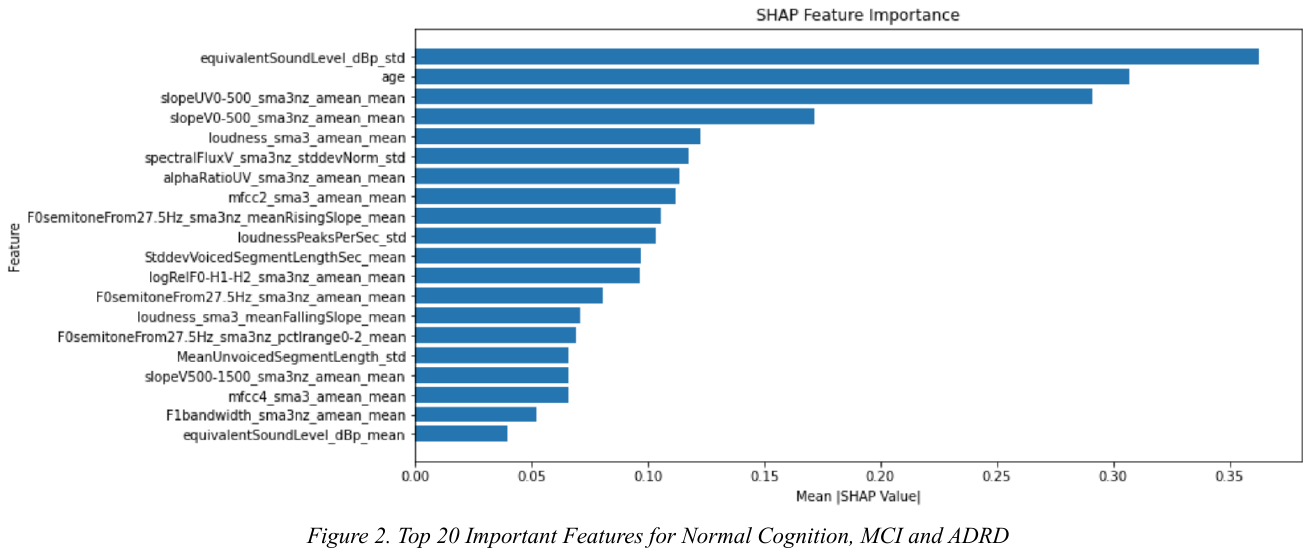}
\caption{Mean absolute SHAP value for the twenty acoustic features with the largest influence on the XGBoost model, pooled across normal cognition, MCI, and ADRD.}
\label{fig:shap-all}
\end{figure}

SHAP values from the XGBoost acoustic model identify which descriptors drive these predictions (Figure~\ref{fig:shap-all}). The largest mean absolute contributions come from the standard deviation of equivalent sound level, age, and the slopes of unvoiced and voiced fundamental frequency in the 0--500~Hz band. Loudness peaks, spectral flux, the alpha ratio, and MFCC statistics follow. Pause and voicing features, including the standard deviation of voiced-segment length and the mean length of unvoiced segments, also rank in the top twenty. Loudness variability, pitch dynamics, spectral shape, and timing therefore carry more discriminative weight than any single speech-rate summary.

\begin{figure}[H]
\centering
\includegraphics[width=0.78\textwidth]{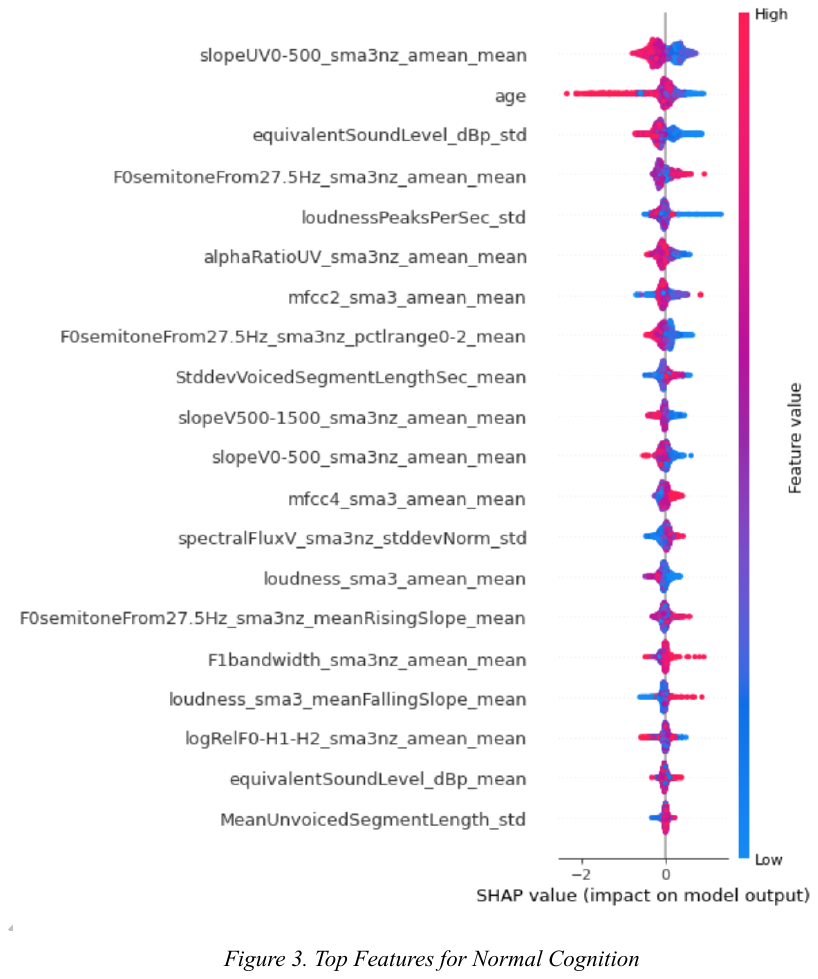}
\caption{SHAP beeswarm for the normal-cognition class. Each point is one recording. Color encodes the feature value.}
\label{fig:shap-control}
\end{figure}

\begin{figure}[H]
\centering
\includegraphics[width=0.78\textwidth]{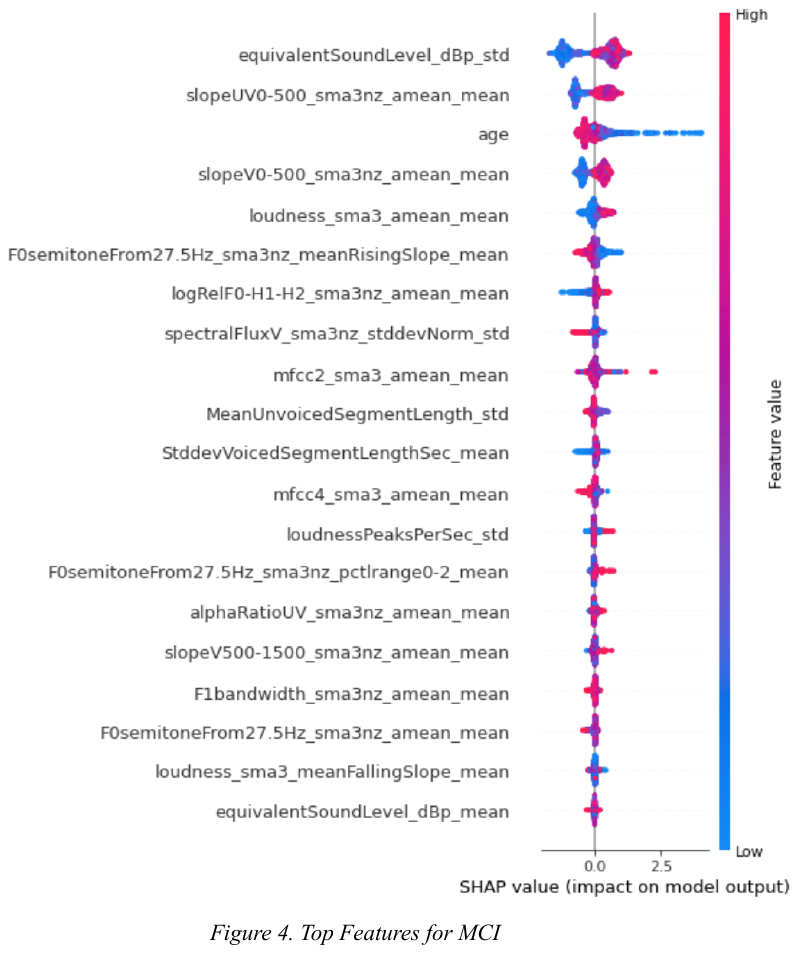}
\caption{SHAP beeswarm for the MCI class.}
\label{fig:shap-mci}
\end{figure}

\begin{figure}[H]
\centering
\includegraphics[width=0.78\textwidth]{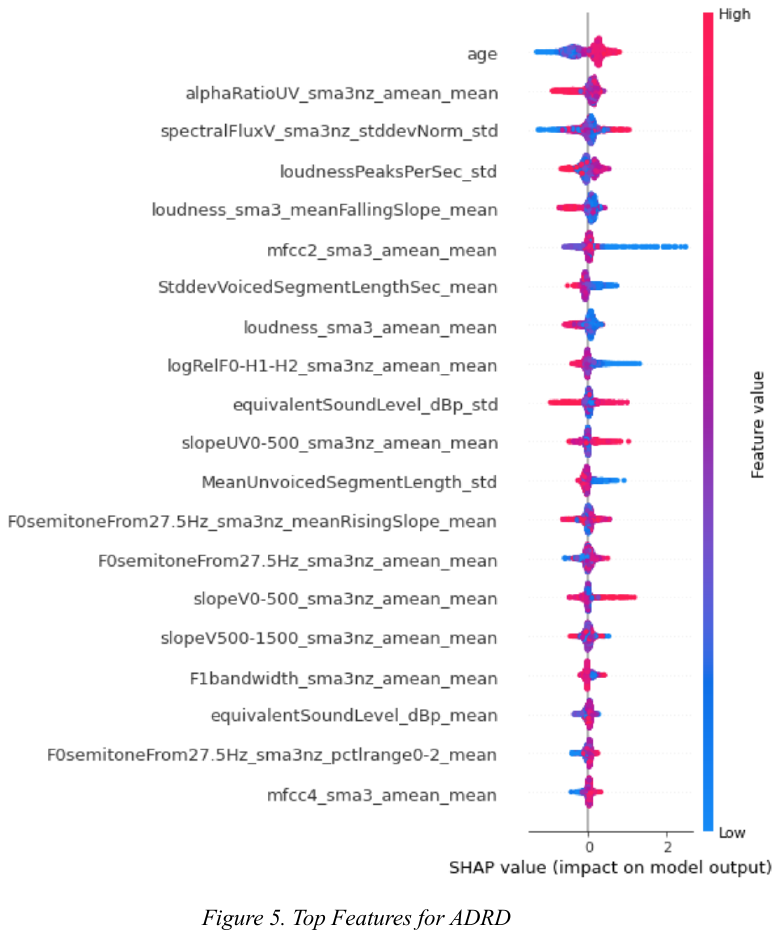}
\caption{SHAP beeswarm for the ADRD class. Age and spectral loudness features rank higher than they do for normal cognition.}
\label{fig:shap-adrd}
\end{figure}

Class-specific beeswarm plots show that the same features do not play the same role in every group. For normal cognition (Figure~\ref{fig:shap-control}), pitch-slope and loudness-peak features dominate, while age contributes less than it does for impaired groups. For MCI (Figure~\ref{fig:shap-mci}), equivalent sound level variability, F0 slope, and unvoiced-segment length become more prominent, which matches longer pauses and less stable intensity as early markers. For ADRD (Figure~\ref{fig:shap-adrd}), age moves to the top of the ranking, and spectral descriptors such as the unvoiced alpha ratio and spectral flux contribute more, alongside loudness and voicing statistics.

These results are consistent with prior acoustic dementia studies, with the additional finding that stacking a contextual language model with a speech embedding lowers log loss beyond either family alone. The acoustic model remains useful because its SHAP profile points to loudness, pitch dynamics, spectral shape, and pause structure rather than to an opaque embedding.

\section{Discussion}

The results support the hypothesis that dementia-related speech changes are reflected in the physical and temporal characteristics of vocal output. Pitch instability, altered energy patterns, more frequent pauses, and greater variability in vocal tract dynamics all provide candidate markers of cognitive impairment. These findings are consistent with earlier work showing that speech prosody, articulation stability, and rhythm are sensitive to neurodegeneration.

An important strength of the present study is its emphasis on interpretable, low-dimensional features. Rather than treating the audio signal as a black box, we derive clinically meaningful descriptors and evaluate whether they retain predictive power. This approach is especially relevant for clinical translation, as interpretability is often critical for healthcare adoption.

At the same time, the study has limitations. The dataset size is moderate, and acoustic markers are influenced by environmental factors, recording conditions, and demographic variation. In addition, speech impairment is not specific to AD; similar patterns can arise from other neurological or psychiatric conditions. Therefore, the model should be understood as a decision-support tool rather than a stand-alone diagnostic instrument. The most promising direction is multimodal fusion, combining acoustic features with linguistic, cognitive, and patient metadata to improve robustness and clinical utility.

From a methodological perspective, future work should incorporate larger and more diverse datasets, external validation across sites, and longitudinal analyses. Such work would allow us to assess whether acoustic biomarkers track disease progression and whether they can detect early cognitive decline before a formal dementia diagnosis is established.

\section{Conclusion}

This study shows that acoustic features extracted from spontaneous speech provide meaningful information for Alzheimer\textquotesingle s disease detection. When paired with standard machine learning methods, these features capture clinically relevant differences in voice production, prosody, and temporal structure. The results are encouraging for the development of non-invasive, scalable screening tools based on speech and audio analysis.

The work demonstrates that speech-based biomarkers are not just a research curiosity but a potentially useful component of digital clinical assessment. By moving beyond purely semantic or linguistic indicators, we open new opportunities for low-cost, early-stage screening and continuous monitoring of cognitive health. Future research should focus on multimodal integration, cross-population validation, and longitudinal tracking to fully realize the value of acoustic biomarkers in dementia care.

\section*{Acknowledgments}

The authors thank the teams and collaborators whose work contributed to the speech and computational paralinguistics literature that informed this study. The present manuscript was prepared as an arXiv-ready research article based on the underlying project materials and associated domain literature.

\bibliographystyle{unsrtnat}
\bibliography{references}

\end{document}